\documentclass[fleqn, 10pt]{wlscirep}
\usepackage[utf8]{inputenc}
\usepackage[T1]{fontenc}
\usepackage{tikz}
\usetikzlibrary{positioning}
\usepackage{multirow}

\title{Towards Efficient PCSEL Design: A Fully AI-driven Approach}

\author[1,2]{Hai Huang}
\author[1,2]{Ziteng Xu}
\author[1,2]{Qi Xin}
\author[1,2,*]{Zhaoyu Zhang}
\affil[1]{School of Science and Engineering, The Chinese University of Hong Kong, Shenzhen, Guangdong 518172, China}
\affil[2]{Guangdong Key Laboratory of Optoelectronic Materials and Chips and Shenzhen Key Lab of Semiconductor Lasers, School of Science and Engineering, The Chinese University of Hong Kong, Shenzhen, Guangdong 518172, China}
\affil[*]{zhangzy@cuhk.edu.cn}

\begin{abstract}
    We present an fully AI-driven design framework for photonic crystals (PhCs), engineered to achieve high efficiency in photonic crystal surface-emitting lasers (PCSELs). By discretizing the PhC structure into a grid, where the edges of the holes are represented by the cross-sections of two-dimensional Gaussian surfaces, we achieve high-degree-of-freedom and fabrication-friendly hole design. Coupled-wave theory (CWT) generates a dataset by evaluating surface-emitting efficiency ($SEE$) and quality factor ($Q$) of PhC designs, while a multi-layered neural network (NN) learns and extracts essential features from these designs. Finally, black-box optimization (BBO) is employed to fine-tune the photonic crystal structure, enabling a fully AI-driven design process. The model achieves high prediction accuracy, with Pearson correlation coefficients of 0.780 for $SEE$ and 0.887 for the log-transformed $Q$. Additionally, we perform Shapley value analysis to identify the most important Fourier coefficients, providing insights into the factors that impact the performance of PCSEL designs. Our work accelerates the design process by over 1,000,000 times compared to traditional FDTD simulations, reducing parameter optimization from two weeks to just one second. Our work speeds up the design process and enables efficient optimization of high-performance PCSELs, driving the development of fully photonic design automation (PDA).
\end{abstract}

\begin{document}
    \flushbottom
    \maketitle
    %
    %
    \thispagestyle{empty}

    \section*{Introduction}

    Photonic crystals (PhCs) are structures with periodic dielectric materials that create photonic band structures, enabling the control of light propagation precisely. These structures have been widely used in various applications, including waveguides, filters, and lasers \cite{butt_Recent_2021,tang_Topological_2022,iadanza_Photonic_2021}. Photonic crystal surface-emitting laser (PCSEL), an important application of PhCs, has attracted significant interest due to their potential across a wide range of applications, including light detection and ranging (LIDAR) systems, laser processing, and optical communications \cite{huang_Unveiling_2025,zhou_Future_2023}. These devices offer the unique advantage of operating in a surface-emitting configuration, which provides several benefits over traditional semiconductor lasers, such as edge-emitting lasers (EELs) and vertical cavity surface-emitting lasers (VCSELs). These
    benefits include high output power, excellent beam quality, an ultra-small divergence angle, and single-mode operation  \cite{noda_Photoniccrystal_2024, ishizaki_Progress_2019,hong_Progress_2022}. Despite these  advantages, traditional methods for designing PhCs often rely on trial-and-error or parameter sweep approaches  \cite{hsieh_Optimal_2024,yoshida_Doublelattice_2019, chang_Photonic_2024,king_Design_2024,chen_Improvement_2021}.
    These methods can be both time-consuming and computationally  expensive, making it difficult to efficiently explore the vast design space of PCSELs.

    AI-driven approaches have gained traction in the field of photonics, as they provide a more efficient alternative for design optimization \cite{alagappan_Leveraging_2022,kudyshev_Machine_2020,chen_POViT_2022}. These methods leverage
    machine learning models to explore complex design spaces and identify optimal
    solutions without the need for exhaustive simulations. By applying these
    techniques, the design process can be significantly accelerated, and the
    computational burden reduced
    \cite{asano_Optimization_2018, liu_Datadriven_2024}. However, most existing
    approaches focus on general optimization rather than providing insights into
    the specific features of PhCs that impact PCSEL, which limits their
    applicability to high-efficiency PCSEL design.

    In this study, we propose a fully AI-driven approach tailored specifically
    for the design of PCSELs. By discretizing the PhC structure into
    grids, our method facilitates the generation of a wide variety of lattice
    configurations, significantly expanding the range of possible designs. One
    of the key features of our approach is the use of Neural Networks (NNs)
    \cite{wallisch_Chapter_2014}, a type of machine learning model well-suited for
    identifying intricate patterns and relationships within complex datasets.
    NNs excel at extracting essential features from the PhC
    structures, such as Fourier coefficients, which serve as a compact
    representation of the structure's key characteristics. By leveraging NNs, we
    are able to capture critical features, such as symmetry and destructive interactions,
    in double-lattice \cite{yoshida_Doublelattice_2019} or triple-lattice
    PhCs \cite{wang_Continuouswave_2024}, that strongly influence laser
    performance.

    Additionally, we employ coupled-wave theory (CWT)
    \cite{liang_Threedimensional_2011, liang_Threedimensional_2012a,
    liang_Threedimensional_2014}
    to evaluate the performance of the PCSEL designs. CWT offers a reliable and efficient
    method for calculating the ratio of surface-emitting to edge-emitting power for
    different resonant modes. By utilizing CWT, we are able to dramatically reduce
    the computational time for each design, enabling rapid evaluation of finite-size
    PCSELs. The synergy between NNs and CWT allows us to accelerate the design and
    optimization process while maintaining high accuracy in efficiency predictions.

    In addition to the performance prediction, we perform Shapley value analysis
    to assess the relative importance of each Fourier coefficient in determining
    the efficiency of PCSEL designs. Shapley values, derived from cooperative
    game theory \cite{shapley_Notes_1951}, offer a quantitative measure of each input
    parameter's impact on the model's output. This analysis reveals the most
    influential features of the PhC structure, providing valuable insights
    into the key factors that drive the performance of PCSELs.

    By combining AI-driven design methods, CWT, and Shapley value analysis,
    our approach not only speeds up the design process but also offers a deeper
    understanding of the structural features that impact PCSEL efficiency. This method
    provides a powerful tool for optimizing PCSELs and offers the potential for greater
    automation in photonic device design. In the following sections, we describe
    the implementation of our approach, present the results from model training
    and testing, and discuss the insights gained from Shapley value analysis.

    \section*{Methods}

    \subsection*{PCSEL simulation}

    To design arbitrary PhCs, we first define a random filling
    factor ($FF$), which determines the ratio of high and low dielectric constant
    materials within the PhC and is expressed as $FF = S_{\mathrm{low\ dielectric}}
    /\left(S_{\mathrm{low\ dielectric}}+S_{\mathrm{high\ dielectric}}\right)$. The unit cell of the PhC is
    discretized into an $n \times n$ grid (for a square lattice). For each grid cell,
    a value $z = G(x, y)$ is generated using the two-dimensional Gaussian surface cross-section fitting method (2D-GCF, Eq.~\ref{eq_2D-GCF_2}).
    This value is compared with a threshold $T_0$ based on the filling factor (FF). If
    $z$ larger than $T_0$, the grid cell is assigned to a
    high dielectric constant material (e.g., GaAs). Otherwise, it is assigned to
    a low dielectric constant material (e.g., air).

    For the numerical analysis of the PCSEL, we apply Coupled-Wave Theory
    (CWT) to analyze a square lattice structure. In this approach, we consider the
    propagation of four basic waves along the $x$- and $y$-directions relative
    to the PhC. The coupling between these waves can be described using
    the coupling matrix $\textbf{C}$, which can be expressed in the following
    form:

    \begin{equation}
        \label{eq1}\textbf{C}=\textbf{C}_{\mathrm{1D}}+\textbf{C}_{\mathrm{rad}}+
        \textbf{C}_{\mathrm{2D}}
    \end{equation}

    where $\textbf{C}_{\mathrm{1D}}$, $\textbf{C}_{\mathrm{rad}}$, and $\textbf{C
    }_{\mathrm{2D}}$ are the coupling matrices for the 1D back-diffraction feedback,
    radiation waves coupling, and 2D diffraction feedback, respectively. They are
    given by the following expressions:

    \begin{equation}
        \label{eq2}\textbf{C}_{\mathrm{1D}}=
        \begin{pmatrix}
            0             & \kappa_{2,0} & 0             & 0            \\
            \kappa_{-2,0} & 0            & 0             & 0            \\
            0             & 0            & 0             & \kappa_{0,2} \\
            0             & 0            & \kappa_{0,-2} & 0
        \end{pmatrix}
    \end{equation}
    \begin{equation}
        \label{eq3}\textbf{C}_{\mathrm{rad}}=
        \begin{pmatrix}
            \zeta_{1,0}^{(1,0)}  & \zeta_{1,0}^{(-1,0)}  & 0                    & 0                     \\
            \zeta_{-1,0}^{(1,0)} & \zeta_{-1,0}^{(-1,0)} & 0                    & 0                     \\
            0                    & 0                     & \zeta_{0,1}^{(0,1)}  & \zeta_{0,1}^{(0,-1)}  \\
            0                    & 0                     & \zeta_{0,-1}^{(0,1)} & \zeta_{0,-1}^{(0,-1)}
        \end{pmatrix}
    \end{equation}
    \begin{equation}
        \label{eq4}\textbf{C}_{\mathrm{2D}}=
        \begin{pmatrix}
            \chi_{y,1,0}^{(1,0)}  & \chi_{y,1,0}^{-1,0}  & \chi_{y,1,0}^{(0,1)}  & \chi_{y,1,0}^{0,-1}  \\
            \chi_{y,-1,0}^{(1,0)} & \chi_{y,-1,0}^{-1,0} & \chi_{y,-1,0}^{(0,1)} & \chi_{y,-1,0}^{0,-1} \\
            \chi_{x,0,1}^{(1,0)}  & \chi_{x,0,1}^{-1,0}  & \chi_{x,0,1}^{(0,1)}  & \chi_{x,0,1}^{0,-1}  \\
            \chi_{x,0,-1}^{(1,0)} & \chi_{x,0,-1}^{-1,0} & \chi_{x,0,-1}^{(0,1)} & \chi_{x,0,-1}^{0,-1}
        \end{pmatrix}
    \end{equation}

    For detailed calculations of the Eq. (\ref{eq2}), Eq. (\ref{eq3}) and Eq. (\ref{eq4}),
    please refer to Ref.~\cite{liang_Threedimensional_2011,liang_Threedimensional_2014}.

    Thus, the we can describe the four basic waves in the PhC as:

    \begin{equation}
        \label{eq5}\left( \delta + i \frac{\alpha_{\mathrm{r}}}{2}\right) \left(
        \begin{array}{cc}
            R_x \\
            S_x \\
            R_y \\
            S_y
        \end{array}
        \right) = \textbf{C}\left(
        \begin{array}{cc}
            R_x \\
            S_x \\
            R_y \\
            S_y
        \end{array}
        \right) + \mathrm{i}\left(
        \begin{array}{cc}
            \partial{R_x}/\partial x  \\
            -\partial{S_x}/\partial x \\
            \partial{R_y}/\partial y  \\
            -\partial{S_y}/\partial y
        \end{array}
        \right)
    \end{equation}

    where $\delta$ is the frequency detuning which can be expressed as
    $\delta = \left({{\beta ^2} - \beta _0^2}\right)/2{\beta _0}\simeq \beta -{\beta _0}
    ={n_{eff}}\left({\omega - {\omega _B}}\right)/c$, with $\omega$ being the frequency
    of the resonant mode and $\omega_{\mathrm{B}}$ being the Bragg frequency.
    The parameter $\alpha_{\mathrm{r}}$ is the the total optical radiation loss
    of the resonant mode. The $R_{x}$, $S_{x}$, $R_{y}$, and $S_{y}$ are the
    amplitudes of the four basic waves. The partial derivatives in the right-hand
    side of Eq. (\ref{eq5}) are the spatial derivatives of the amplitudes of the
    four basic waves.

    Applying the boundary conditions
    $R_{x}(0, y) = S_{x}(L, y) = R_{y}(x, 0) = S_{y}(x, L) = 0$ for a finite-size
    $L \times L$ square-lattice PhC, we can solve Eq. (\ref{eq5})
    to obtain the spatial distribution of four basic waves. We can then calculate
    the optical power of the surface-emitting and edge-emitting, and derive the
    efficiency of the PCSEL. The optical power of all stimulated emission,
    surface-emitting and emitting can be expressed as:

    \begin{equation}
        \label{eq6}P_{\mathrm{stim}}=\alpha_{\mathrm{r}}\iint_{0}^{L}(|R_{x}|^{2}
        +|S_{x}|^{2}+ |R_{y}|^{2}+|S_{y}|^{2})\mathrm{d}x\mathrm{d}y
    \end{equation}
    \begin{equation}
        \label{eq7}P_{\mathrm{edge}}=\int_{0}^{L}(|R_{x}|^{2}+|S_{x}|^{2})|_{x=0,L}
        \mathrm{d}y+\int_{0}^{L}(|R_{y}|^{2}+|S_{y}|^{2})|_{y=0,L}\mathrm{d}x
    \end{equation}
    \begin{equation}
        \label{eq8}P_{\mathrm{surface}}=2\Im(\kappa_{v})\iint_{0}^{L}(|\xi_{-1,0}
        R_{x}+\xi_{1,0}S_{x}|^{2}+| \xi_{0,-1}R_{y}+\xi_{0,1}S_{y}|^{2})\mathrm{d}
        x\mathrm{d}y
    \end{equation}

    Note that, in Eq. (\ref{eq8}), $\kappa_{v}$ is a term associated with
    radiation waves coupling and $\xi$ is Fourier coefficient (seeing Ref.
    \cite{liang_Threedimensional_2012a}).

    To evaluate the efficiency of the designed PCSEL, we consider the ratio of
    surface-emitting power to the stimulated emission power as surface-emitting
    efficiency ($SEE$). Since we are designing a surface-emitting laser, a
    higher ratio of $SEE = P_{\mathrm{surface}}/ P_{\mathrm{stim}}$ indicates
    higher efficiency of the laser. Specifically, the efficiency is enhanced
    when the surface-emitting power is maximized relative to the total
    stimulated emission, which leads to greater power extraction through the surface
    of the device. This ratio can be used as a key performance indicator for
    optimizing the PhC structure for high-efficiency laser
    operation.

    In addition to surface-emitting efficiency, the value of the quality factor $Q$
    is also an important metric for evaluating the performance of PCSELs. The
    quality factor is defined as the ratio of the energy stored in the cavity to
    the energy lost per cycle, and is given by $Q = \frac{2\pi / a}{\alpha_{\mathrm{r}}}$,
    where $a$ is the lattice constant of the PhC. A high quality factor
    indicates low cavity loss and high energy storage, which are essential for
    low-threshold lasing. Therefore, the quality factor is also a key result
    that we need to extract from the model to ensure the laser operates at its
    optimal performance.To handle the large variation in the value of $Q$ (ranging
    from $10^{3}$ to $10^{5}$), we trained the model to predict $\log{Q}$ instead
    of $Q$ directly. This transformation reduces the dynamic range of the target
    values, thereby stabilizing the training process and improving convergence.
    After prediction, the final $Q$ value is obtained by taking the exponential of
    the model output. This approach ensures that the model captures the
    underlying pattern more effectively while minimizing sensitivity to extreme
    variations in $Q$.

    Although the partial differential equation (PDE) analysis based on CWT is
    computationally fast (requiring only a few seconds), it becomes a significant
    bottleneck when performing large-scale parameter optimization and analysis.
    Consequently, to address this challenge and speed up the evaluation process,
    we propose the use of neural networks to predict $SEE$ and $\log{Q}$, bypassing
    the need for extensive numerical computations. By training a neural network
    model on the simulated data, we can achieve rapid predictions of $SEE$ and $\log
    {Q}$, significantly reducing the time required for optimization.

    \subsection*{Neural network architecture}

\begin{figure}[htbp]
        \centering
        \includegraphics[width=12cm]{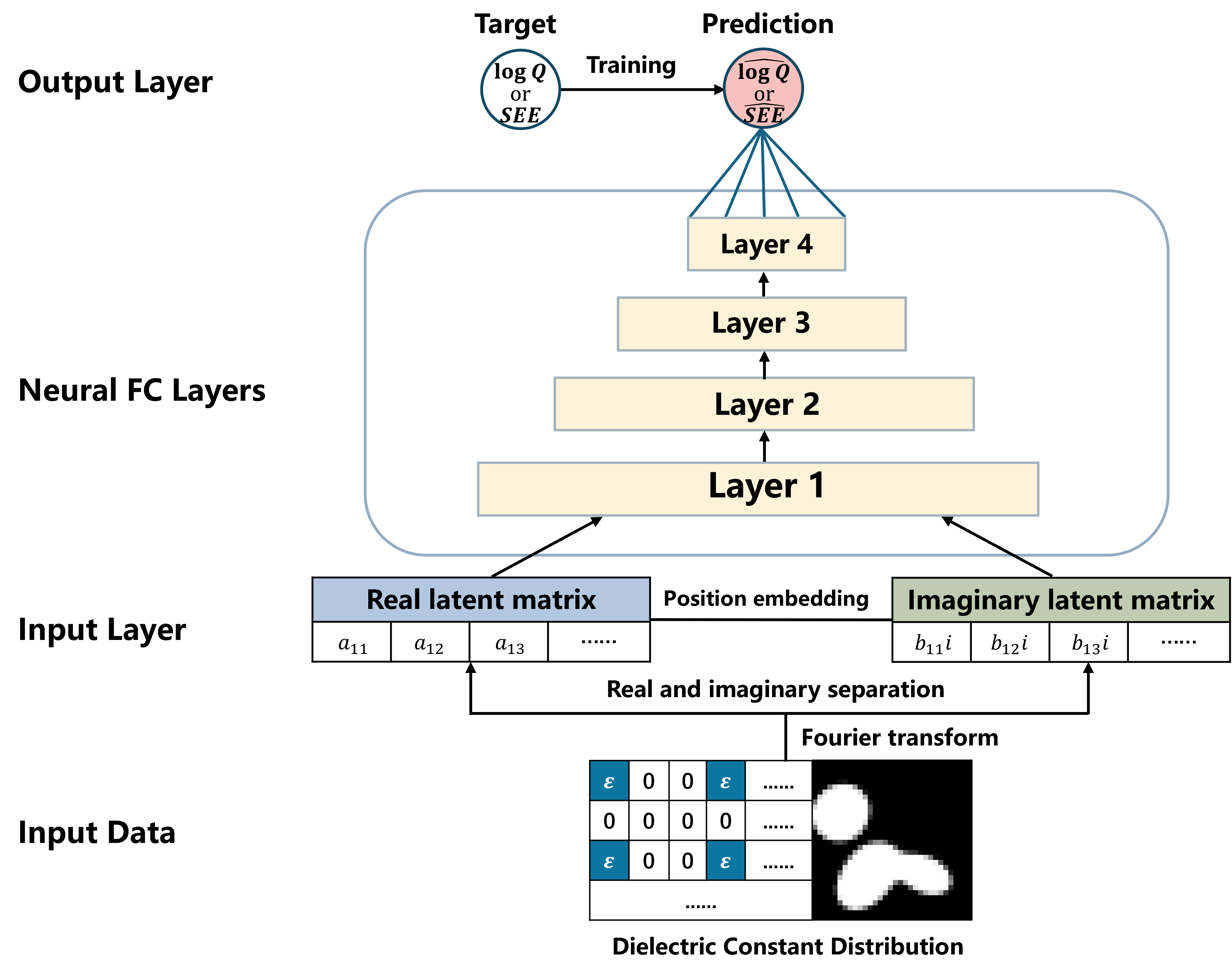}
        \caption{The architecture of the neural network used for predicting the
        efficiency of PCSELs. The real latent matrix $a$ corresponds to the Fourier
        coefficients $\xi^{R}_{m,n}$, while the imaginary latent matrix $b$
        represents the Fourier coefficients $\xi^{I}_{m,n}$. The network consists
        of four layers (Layer 1 to Layer 4), with each layer learning to map the
        input coefficients to the final prediction of either $\log{Q}$ or $SEE$.}
        \label{fig:NN}
    \end{figure}

    We developed a fully connected neural network (shown in Figure~\ref{fig:NN}),
    to predict $SEE$ and $\log{Q}$ of PCSEL designs. The model begins by 1024 input
    grid, which represents the real and imaginary parts of the Fourier
    coefficients ($\xi^{X}_{m,n}$, where $X \in \{R, I\}$) of the discretized PhC structure. A learnable positional embedding \cite{wang_What_2020} ($P
    E^{R}_{m,n}$ and $PE^{I}_{m,n}$) is then added to this flattened input to
    help capture spatial information. We set $PE^{R}_{m,n}= PE^{I}_{m,n}$ to ensure
    that the real and imaginary parts of the Fourier coefficients have the same positional
    embedding. For simplicity and to reduce redundancy, we only retain one of the
    pairings of $\xi^{X}_{m,n}$ and $\xi^{X}_{-m,-n}$ in our analysis. This is
    because these coefficients are complex conjugates of each other and
    represent the same input. By removing the conjugate pair, we avoid
    duplicating the information, ensuring more efficient processing without loss
    of relevant data.

    The network consists of four fully connected layers with 1024, 512, 256, and
    64 neurons, respectively. Given that the matrix is complex, it is logical to
    concatenate real and Imaginary latent matrix, using a Multi-Layer Perceptron
    (MLP) to learn the interaction. In this context, NNs can be defined as
    \begin{align}
        \label{eq9}z_{1} & = W_{0}^{T}x + b_{0},        \\
        \phi(z_{1})      & = a(W_{1}^{T}z_{1}+ b_{1}) , \\
                         & \dots ,                      \\
        \phi(z_{4})      & = a(W_{4}^{T}z_{4}+ b_{4}) , \\
        y                & = \sigma(\phi(z_{4})),
    \end{align}
    where $W_{x}$ and $b_{x}$ denoted as layer weights, $a$ is Parametric ReLU (PReLU)
    activations and $\sigma$ is a sigmoid function in training SEE to ensure its
    value is constrained within the range $(0,1)$ and $\sigma$ is identity transformation
    in training $\log Q$.

    This architecture is designed to effectively learn the mapping between the
    grid-based representation of the PhC and its performance, enabling
    accurate prediction of the laser's performance.

    \subsection*{Generation of dataset}

    The dataset for training and testing was generated using the two-dimensional Gaussian surface cross-section fitting method (2D-GCF). In this approach, random parameters are assigned to define the characteristics of multiple two-dimensional Gaussian distributions. Specifically, for each Gaussian distribution, random values are generated for the amplitude $A$, center positions $(x_0, y_0)$, standard deviations $(\sigma_x, \sigma_y)$, and orientation angles $(\theta)$. These individual Gaussian distributions are then summed to form a composite function $G(x, y)$. Finally, the contour where $G(x, y)$ equals a specific threshold value $T_0$ is extracted to define the edges of the holes in the PhC structure. The resulting clear hole contours are pixelated into a $32 \times 32$ grid map (dielectric array) for further processing. Due to the inherent smoothness of Gaussian functions, the designed hole contours are guaranteed to be smooth, which provides significant convenience for subsequent fabrication processes. Moreover, this randomization and summation process ensures a diverse and comprehensive dataset, capturing a wide range of possible configurations for robust model training and evaluation. In this study, the authors fixed the number of Gaussian distributions to three.

    \begin{equation}
        \label{eq_2D-GCF_1}g_i(x, y) = A_i \cdot \exp\left(-\frac{(x - x_{0,i})^2}{2\sigma_{x,i}^2} - \frac{(y - y_{0,i})^2}{2\sigma_{y,i}^2}\right),     i \in \mathbb{Z}^+
    \end{equation}

    \begin{equation}
        \label{eq_2D-GCF_2}G(x, y) = \sum_{i=1}^{n} g_i(x, y)
    \end{equation}

    \begin{equation}
        \label{eq_2D-GCF_3}G(x, y) = T_0
    \end{equation}

    The simulation function is subsequently called with the generated dielectric
    array and a solver instance. This function performs optical simulations
    based on coupled-wave theory (CWT) and finite element method (FEM) analysis,
    and returns key performance metrics such as the $SEE$ and quality factor $Q$,
    and other relevant parameters. We adopt the epitaxial structure outlined in Table~\ref{tab:epitaxy-structure},
    with the Bragg wavelength fixed at 980 nm. To ensure consistency with the finite-size
    PCSELs, the PhC's lateral dimension $L$ is chosen to be as
    close as possible to 200 $\mu m$, while maintaining an integer multiple of the
    crystal period. The finite-size PhC are discretized to a
    $17 \times 17$ grid for FEM calculation. The results
    are then written to a NPZ file (see supplementary information), which serves
    as the dataset for further model training and testing. This iterative process
    continues until an external stop signal is detected, thereby building a
    comprehensive dataset of PhC designs. Finally, our dataset
    consists of 200k simulation results, ensuring a diverse and representative set
    of samples for training robust machine learning models.

    \begin{table}[htbp]
        \caption{Epitaxial structure of the PCSEL in this work.}
        \label{tab:epitaxy-structure}
        \centering
        \begin{tabular}{cccc}
            \hline
            Layer                    & Material      & Thickness ($\mu m$) & Refractive Index \\
            \hline
            Photonic Crystal         & p-GaAs/Air    & 0.35                & 3.4826/1         \\
            Waveguide                & p-GaAs        & 0.08                & 3.4826           \\
            Electron Blocking Layers & p-AlGaAs      & 0.025               & 3.2806           \\
            Active Region            & InGaAs/AlGaAs & 0.116               & 3.3944           \\
            n-cladding               & n-AlGaAs      & 2.11                & 3.2441           \\
            n-substrate              & n-GaAs        & -                   & 3.4826           \\
            \hline
        \end{tabular}
    \end{table}

    The epitaxial structure used in this work, as shown in Table~\ref{tab:epitaxy-structure},
    serves as an example for the design of the PCSEL. The choice of epitaxial layers
    mainly impacts the Green's function and the confinement factor of the photonic
    crystal ($\Gamma_{\mathrm{PhC}}$) within the framework of CWT. However, since
    the primary focus of this work is on the design of the PhC structure
    itself, the specific epitaxial layers chosen do not significantly affect the
    overall performance. Furthermore, the methodology proposed here can be
    easily adapted to other epitaxial structures, making the design approach flexible
    for various PCSELs epitaxial structures.

    \section*{Results and discussions}

    \subsection*{Model training and testing}

\begin{figure}[htbp]
        \centering
        \begin{tikzpicture}
            \scope[nodes={inner sep=0,outer sep=0}]
            \node[anchor=south east]
                (a)
                {\includegraphics[width=4cm]{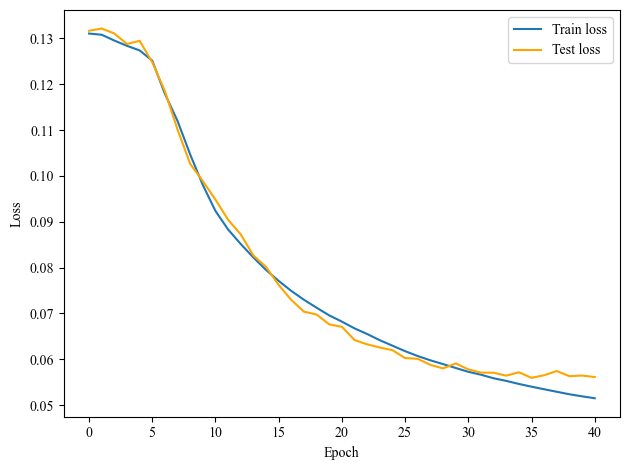}};
            \node[right=3mm of a.north east, anchor=north west]
                (b)
                {\includegraphics[width=4cm]{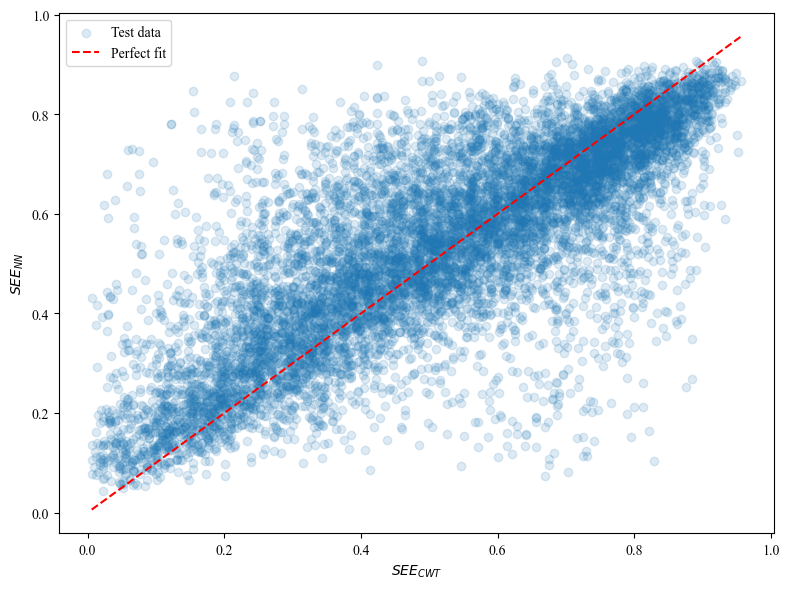}};
            \node[right=3mm of b.north east, anchor=north west]
                (c)
                {\includegraphics[width=4cm]{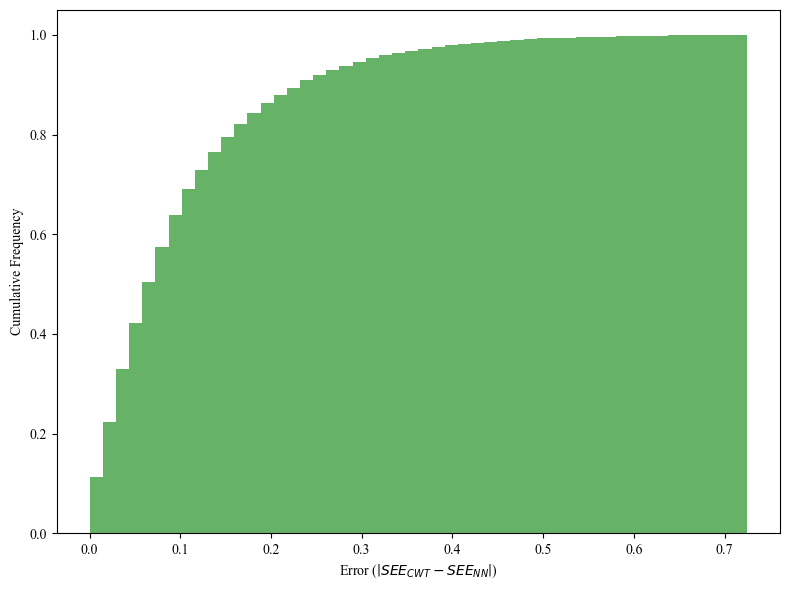}};
            \node[below=1mm of a.south west, anchor=north west]
                (d)
                {\includegraphics[width=4cm]{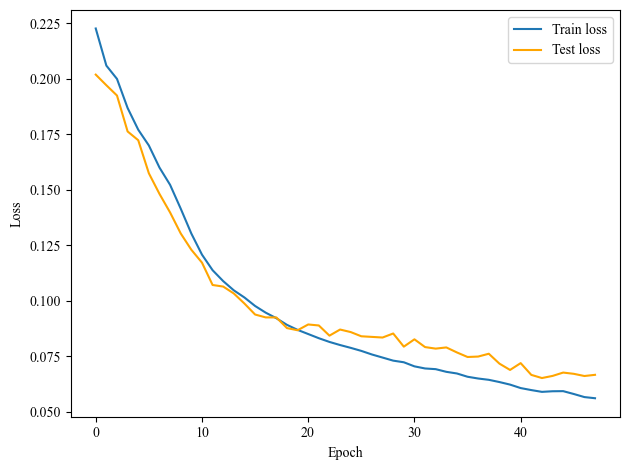}};
            \node[below=1mm of b.south west, anchor=north west]
                (e)
                {\includegraphics[width=4cm]{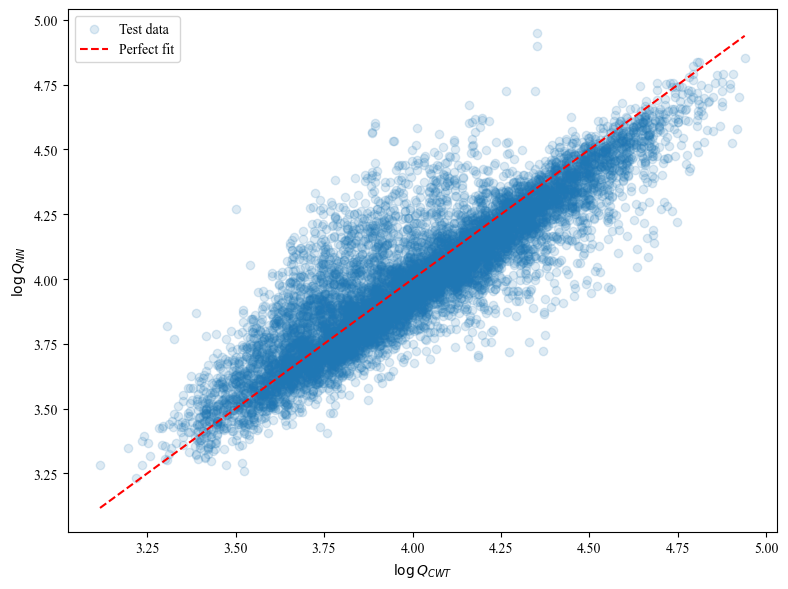}};
            \node[below=1mm of c.south west, anchor=north west]
                (f)
                {\includegraphics[width=4cm]{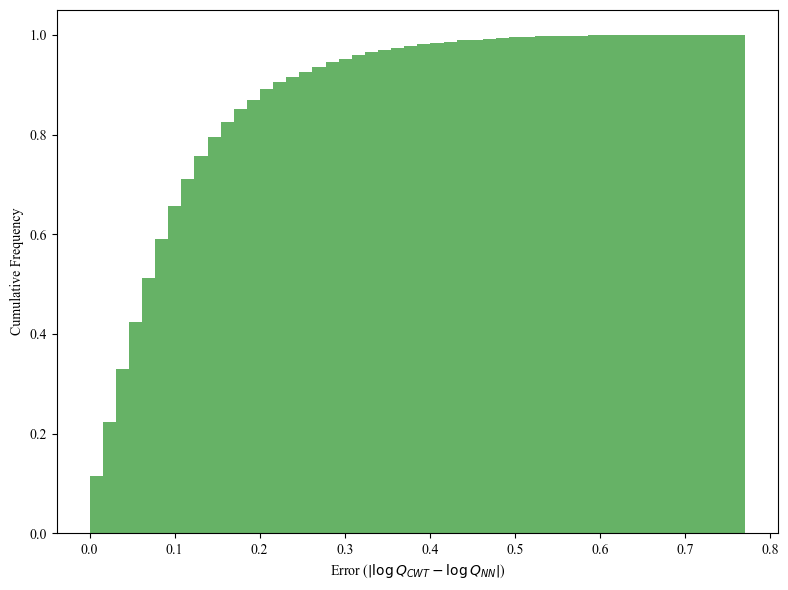}};
            \endscope \foreach \n in {a,b,c,d,e,f}
            { \node[right=2mm of \n.north west, anchor=north east] {(\n)}; }
        \end{tikzpicture}
        \caption{Training results for $SEE$ and $Q$ models. (a) and (d) show the
        learning curves for surface-emitting efficiency ($SEE$) and log-transformed
        quality factor ($\log{Q}$) during training, respectively. (b) and (e) display
        the corresponding test results for $SEE$ and $\log{Q}$, while (c) and (f)
        present the error distribution for the training results of $SEE$ and
        $\log{Q}$, respectively.}
        \label{fig:training-results}
    \end{figure}

    We trained a fully connected neural network to predict key performance metrics
    of PCSEL designs based on the generated
    dataset. The dataset was split into 190k samples for training and 10k samples
    for testing. The model was trained using the Adam optimizer and training
    process was run for up to 100 epochs, with early stopping based on the validation
    loss to prevent overfitting (shown in Figure~\ref{fig:training-results} (a)
    and (d)). Train epoch finally stopped at about 45 epochs. We evaluated the model's
    performance on the test set using the Pearson correlation coefficient as the
    primary metric. Pearson's correlation coefficient is used to evaluate the
    linear relationship between predicted and actual values, providing insight into
    the model's predictive accuracy. The results are summarized in Table~\ref{tab:performance-metrics},
    which shows a high correlation between the predicted and actually simulated efficiency
    values. In Figure~\ref{fig:training-results} (b) and (e), we present the test
    results for $SEE$ and $\log{Q}$, respectively. These results demonstrate that
    the model effectively captures the key performance characteristics of PCSEL designs,
    with relatively high prediction accuracy. The error distribution for the training
    results of $SEE$ and $\log{Q}$ is shown in Figure~\ref{fig:training-results}
    (c) and (f), respectively. About 80\% of the predictions fall within a range
    of $\pm 0.14$ for $SEE$ and $\pm 0.15$ for $\log{Q}$, indicating that the model
    produces accurate estimates with minimal deviation from the simulated values.

    \begin{table}[htbp]
        \caption{Performance metrics of the NN model on the test dataset for
        predicting the quality factor $\log{Q}$ and surface-emitting efficiency
        $SEE$.}
        \label{tab:performance-metrics}
        \centering
        \begin{tabular}{cccc}
            \hline
            Metric    & Pearson Correlation Coefficient (and p-value) & RMSE  & $R^{2}$ \\
            \hline
            $SEE$     & 0.780 (p < 0.001)                             & 0.144 & 0.595   \\
            $\log{Q}$ & 0.887 (p < 0.001)                             & 0.140 & 0.784   \\
            \hline
        \end{tabular}
    \end{table}

    As shown in Table~\ref{tab:performance-metrics} these results demonstrate that
    the neural network model effectively captures the key performance characteristics
    of PCSEL designs. The Pearson correlation coefficients which means the
    linear relationship between predicted and actual values, are 0.780 for $SEE$
    and 0.887 for $\log{Q}$. These high correlation values indicate that the
    model has strong predictive capability. The relatively low RMSE (Root of
    Mean Square Error) values further suggest that the model produces accurate estimates
    with a minimal deviation from the simulated values. We also calculated the
    coefficient of determination ($R^{2}$) to evaluate the proportion of
    variance in the data that is explained by the model. A $R^{2}$ value closer to
    1 indicates that the model explains a larger portion of the variance in the data,
    confirming its predictive power. In our results, the $R^{2}$ values of 0.595
    for $SEE$ and 0.784 for $\log{Q}$ confirm that the model explains a
    substantial portion of the variance in the data. These findings highlight the
    potential of fully AI-driven approaches for accelerating the optimization of PCSEL
    designs, reducing reliance on computationally expensive simulations while
    maintaining high prediction accuracy.

    \subsection*{Global optimization of PhC design}

    After establishing the predictive capabilities of our trained neural network
    models for $SEE$ and $Q$, we
    leverage these models to search for a PhC design that minimizes
    $\left|SEE_{Target}-SEE\right|$ while ensuring $Q > 10^{4}$ to maintain a
    low operational threshold.

    To achieve this, we define a loss function:
    \begin{equation}
        \mathcal{L}= \left|SEE_{Target}-SEE\right| + \lambda \max(0, 10^{4}- Q),
    \end{equation}
    where $SEE_{Target}$ is the desired surface-emitting efficiency, and $\lambda$
    (is setted to 5e-5) is a penalty coefficient that enforces the quality factor
    constraint. A larger penalty is applied when $Q$ falls below the required
    threshold, ensuring that solutions with desired $SEE$ and sufficiently large
    $Q$ are favored.

    We employ a global black-box optimization (BBO) algorithm, specifically the DIRECT
    \cite{gablonsky_Locallybiased_2001,jones_Lipschitzian_1993} (DIviding RECTangles)
    method, to efficiently explore the parameter space. The optimization
    variables include the positions, sizes, and orientations of the holes in the
    PhC unit cell, as well as the filling factor. By utilizing the
    trained models for $SEE$ and $Q$, the optimization process is significantly accelerated,
    as each evaluation bypasses the need for relatively computationally
    expensive coupled-wave theory (CWT) simulations. To validate the optimal design
    obtained via the model-driven optimization, we perform full physical
    simulations using the CWT and FEM solvers. A comparison between predicted and
    simulated results further confirms the reliability of the neural network models
    in guiding the optimization process. The optimized structure is summarized
    in Table~\ref{tab:optimized-structure}. The final validation results demonstrate
    strong agreement with our model predictions. The entire optimization process
    involved over 16,000 function evaluations, with a total runtime of approximately
    only 2.2 minutes. This significantly accelerates the design process by over 1,000,000 times compared to traditional FDTD simulations, which usually cost several hours per finite PCSEL simulation on standard workstation computers. This demonstrates the effectiveness of combining machine learning-based models with BBO algorithm for rapid and accurate PhC design.

    \begin{table}[htbp]
        \centering
        \caption{Optimized photonic crystal structure for surface-emitting
        efficiency ($SEE$) and quality factor ($Q$). The dielectric constant distributions
        correspond to different $SEE_{Target}$ values, and both predicted and
        validated performance metrics are listed.}
        \label{tab:optimized-structure}
        \begin{tabular}{c|c|c|c|c|c|c}
            \hline
            $SEE_{Target}$                                        & \multicolumn{2}{c|}{$1.000$}                                                      & \multicolumn{2}{c|}{$0.500$}                                                      & \multicolumn{2}{c}{$0.000$}                                                      \\
            \hline
            \raisebox{3\height}{Dielectric constant distribution} & \multicolumn{2}{c|}{\raisebox{-0.15cm}{\includegraphics[width=2cm]{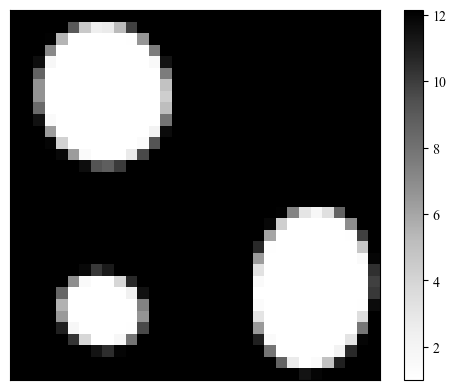}}} & \multicolumn{2}{c|}{\raisebox{-0.15cm}{\includegraphics[width=2cm]{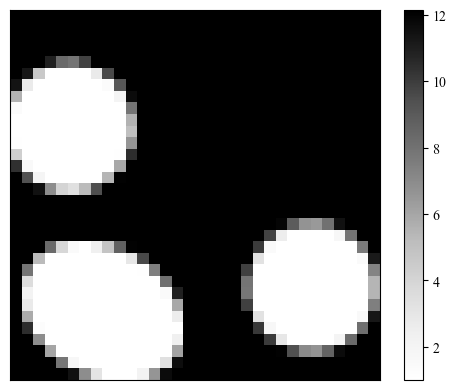}}} & \multicolumn{2}{c}{\raisebox{-0.15cm}{\includegraphics[width=2cm]{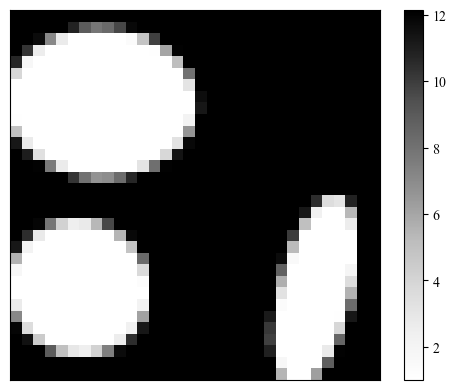}}} \\
            \hline
            Results                                               & $SEE$                                                                             & $Q$                                                                               & $SEE$                                                                           & $Q$     & $SEE$ & $Q$     \\
            \hline
            Predicted                                             & 0.889                                                                             & 10001.9                                                                           & 0.500                                                                           & 18643.2 & 0.034 & 13508.0 \\
            \hline
            Validated                                             & 0.867                                                                             & 9524.2                                                                            & 0.477                                                                           & 19015.2 & 0.017 & 16666.2 \\
            \hline
            Number of calls                                       & \multicolumn{2}{c|}{16015}                                                        & \multicolumn{2}{c|}{16135}                                                        & \multicolumn{2}{c}{16007}                                                        \\
            \hline
            Total runtime                                         & \multicolumn{2}{c|}{130.4s}                                                       & \multicolumn{2}{c|}{143.1s}                                                       & \multicolumn{2}{c}{131.2s}                                                       \\
            \hline
        \end{tabular}
    \end{table}

    This approach demonstrates the potential of machine learning models not only
    in predicting PhC performance but also in guiding global
    optimization for high-efficiency laser design.

    \subsection*{Shaply values analysis}

    We employed Shapley values, a game-theoretic approach for interpreting
    machine learning models \cite{shapley_Notes_1951}, to evaluate the influence
    of each input on the neural network model's predictions. Since our input
    features are represented as the real and imaginary parts of the Fourier
    series coefficients of the PhC structure, we computed Shapley values
    for each component separately to quantify their contribution to the predicted
    $SEE$ and $Q$. The Shapley value analysis was implemented using the SHAP
    library \cite{lundberg_Unified_2017}. This approach helps us understand which
    Fourier coefficients most significantly affect the surface-emitting efficiency
    and quality factor, providing insight into the design of PCSELs.

\begin{figure}[htbp]
        \centering
        \begin{tikzpicture}
            \scope[nodes={inner sep=0,outer sep=0}]
            \node[anchor=south east]
                (a)
                {\includegraphics[width=6.3cm]{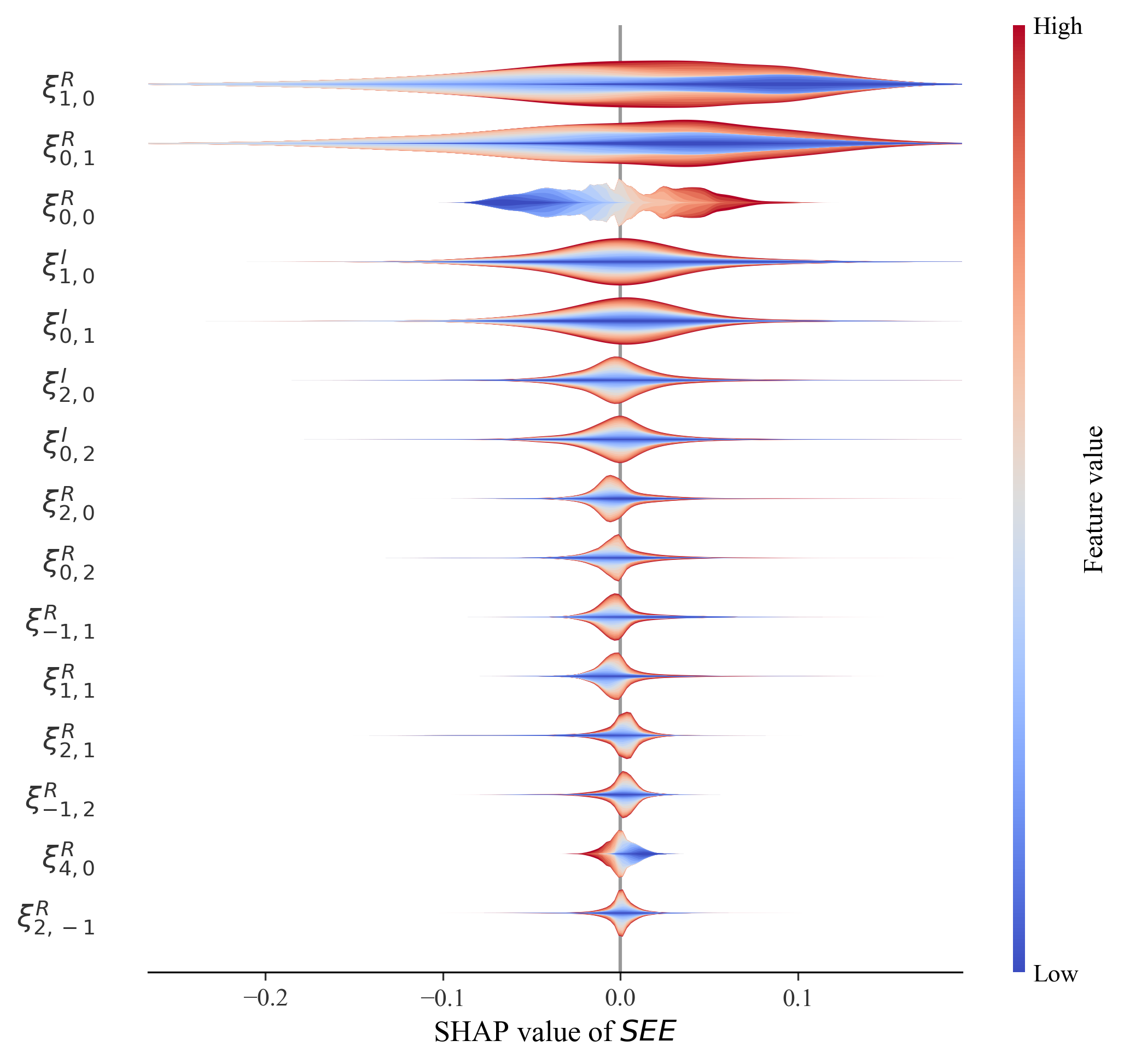}};
            \node[right=3mm of a.north east, anchor=north west]
                (b)
                {\includegraphics[width=6.3cm]{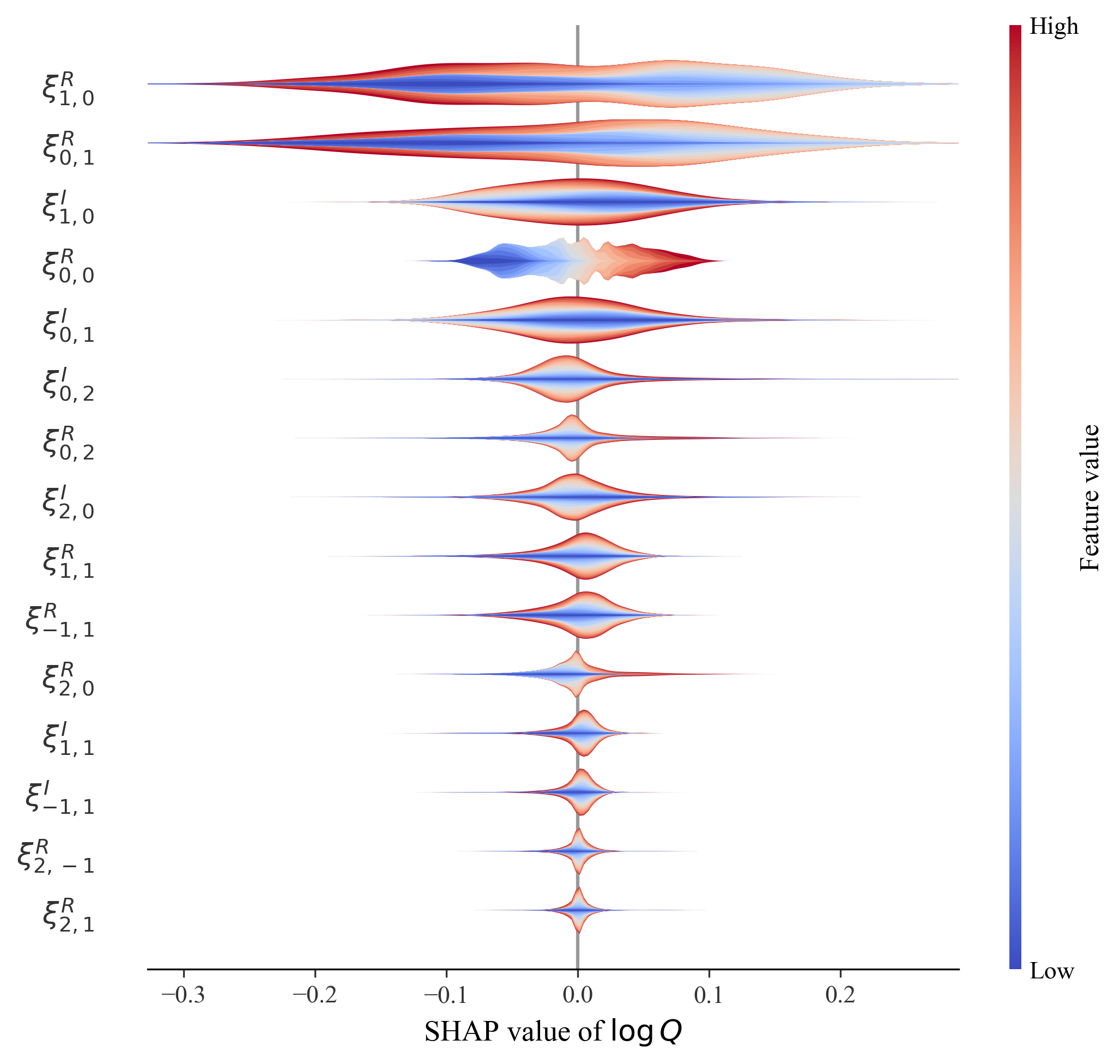}};
            \endscope \foreach \n in {a,b}
            { \node[right=2mm of \n.north west, anchor=north east] {(\n)}; }
        \end{tikzpicture}
        \caption{SHAP (Shapley Additive Explanations) values for the top 15 most
        influential Fourier coefficients in predicting the $SEE$ (a) and
        $\log{Q}$ (b). Each row represents a Fourier coefficient, with the SHAP
        values shown along the x-axis. The color scale indicates the magnitude of
        the feature values, where red represents high values and blue represents
        low values. The violin plots show the distribution of SHAP values for each
        feature, highlighting its impact on the predicted $SEE$ or $\log{Q}$. Positive
        SHAP values correspond to features that increase $SEE$ or $\log{Q}$,
        while negative values correspond to features that decrease $SEE$ or
        $\log{Q}$.}
        \label{fig:shapley-values}
    \end{figure}

    To gain the relative impacts of different input features in determining the
    performance of PCSEL designs, we conducted a Shapley value analysis (on 500
    randomly selected samples from the test dataset). The results of this
    analysis are shown in Figure~\ref{fig:shapley-values}, which illustrates the
    contribution of each Fourier coefficient to the predicted $SEE$ and $\log{Q}$.
    The Shapley values provide a quantitative measure of the impact of each
    input parameter on the model's output, revealing the most influential features
    in the PhC structure.

    For $SEE$, the Fourier coefficients $\xi^{R}_{1,0}$ and $\xi^{R}_{0,1}$ have
    the most significant influence. Higher deviation values of these
    coefficients (further from the mean) lead to an increase in $SEE$, while values
    closer to the mean result in a decrease in $SEE$. For $\log{Q}$, the trend is
    reversed. Higher deviation values of $\xi^{R}_{1,0}$ and $\xi^{R}_{0,1}$
    lead to a decrease in $\log{Q}$, while values closer to the mean result in an
    increase in $\log{Q}$. This behavior is due to the direct relationship
    between $\xi^{R}_{1,0}$ and $\xi^{R}_{0,1}$ with radiation coupling. When these
    coefficients deviate more from the mean, they increase the radiation loss, thereby
    decreasing $\log{Q}$, while more moderate values allow for better
    confinement and lower radiation loss.

    The $\xi^{R}_{0,0}$, which corresponds to the average dielectric constant of
    the PhC layer, also has a notable influence on both $SEE$ and
    $\log{Q}$. $\xi^{R}_{0,0}$ shows a generally positive correlation with both $S
    EE$ and $\log{Q}$. This suggests that higher values of $\xi^{R}_{0,0}$ lead
    to improved performance in both surface-emitting efficiency and the quality
    factor. We speculate that this is because $\xi^{R}_{0,0}$ directly influences
    the confinement factor ($\Gamma_{\mathrm{PhC}}$) within the PhC,
    which in turn affects both the efficiency of surface-emitting power
    extraction and the quality factor. A higher $\xi^{R}_{0,0}$ means a better confinement
    of light within the structure, leading to enhanced optical performance, including
    higher $SEE$ and $\log{Q}$ values.

    We summarized the impact of part of different Fourier coefficients on $SEE$ and
    $\log{Q}$ in Table~\ref{tab:fourier-coefficients-impact}. Other Fourier
    coefficients also contribute to the results, but their influence is less straightforward,
    indicating that there are complex interactions between them.

    \begin{table}[htbp]
        \caption{Summary of the influence of Fourier coefficients on surface-emitting
        efficiency ($SEE$) and the quality factor ($\log{Q}$). The symbols "+" and
        "-" indicate positive and negative correlations, respectively. The number
        of "+" or "-" symbols represents the magnitude of the influence, with
        more symbols indicating a stronger effect on $SEE$ and $\log{Q}$. "\textbackslash"
        indicates that the feature does not has direct impact on the corresponding
        result.}
        \label{tab:fourier-coefficients-impact}
        \centering
        \begin{tabular}{|c|c|c|}
            \hline
            Feature                                                                                                           & Impact on $SEE$ & Impact on $\log{Q}$ \\
            \hline
            $\left|\xi^{R}_{1,0}-\overline{\xi^{R}_{1,0}}\right|$ and $\left|\xi^{R}_{0,1}-\overline{\xi^{R}_{0,1}}\right|$   & $+++$           & $---$               \\
            \hline
            $\xi^{R}_{0,0}$                                                                                                   & $++$            & $++$                \\
            \hline
            $\xi^{R}_{1,1}$                                                                                                   & $+$             & \textbackslash      \\
            \hline
            $\left|\xi^{R}_{1,1}-\overline{\xi^{R}_{1,1}}\right|$ and $\left|\xi^{R}_{-1,1}-\overline{\xi^{R}_{-1,1}}\right|$ & \textbackslash  & $-$                 \\
            \hline
            $\xi^{R}_{2,0}$ and $\xi^{R}_{0,2}$                                                                               & \textbackslash  & $+$                 \\
            \hline
            $\xi^{R}_{4,0}$                                                                                                   & $-$             & \textbackslash      \\
            \hline
        \end{tabular}
    \end{table}

    \section*{Conclusion}

    In this work, we developed a PCSELs simulation framework based on CWT to analyze and optimize the performance of PCSELs. By systematically modeling the interaction of optical waves
    within a structured photonic lattice, we derived key performance metrics,
    including surface-emitting efficiency ($SEE$) and the quality factor ($Q$). To
    further enhance the design process, we implemented a neural network model
    trained on a large dataset of simulated PCSEL structures. of the PhC,
    enabling rapid optimization of device performance.

    Additionally, we employed Shapley value analysis to interpret the contributions
    of different Fourier components, providing insights into the structural
    features that most significantly impact laser efficiency. Our results demonstrate
    that machine learning can serve as a powerful tool for guiding PhC
    design, accelerating the discovery of high-performance PCSEL configurations.
    The methodology proposed in this study is flexible and can be extended to different
    epitaxial structures and laser designs, paving the way for future
    advancements in high-efficiency surface-emitting lasers.

    \bibliography{sample}

    \section*{Acknowledgements}

    This research was supported by the National Natural Science Foundation of China (NSFC) under Grant No. 62174144, the Shenzhen Science and Technology Program under Grants No. JCYJ20220818102214030, No. KJZD20230923115114027, No. JSGG20220831093602004, No. KJZD20240903095602004, the Guangdong Key Laboratory of Optoelectronic Materials and Chips under Grant No. 2022KSYS014, the Shenzhen Key Laboratory Project under Grant No. ZDSYS201603311644527, the Longgang Key Laboratory Project under Grants No. ZSYS2017003 and No. LGKCZSYS2018000015, and the Innovation Program for Quantum Science and Technology (Grant No. 2021ZD0300701), Hefei National Laboratory, Hefei 230088, China.
    The authors would thank Ms. Floria Chen and Ms. Chuanyi Yang for their mentorship and revision tips. The authors would also thank Mr. Sixuan Mao for suggestions on frameworks of artificial neural networks.

    \section*{Author contributions statement}

    H.H. conceived the experiment and conducted the majority of the manuscript writing. H.H., Z.X. and Q.X. conducted the experiment and analyzed the results. Z.X. wrote the section on the principles of the neural network and prepared Figure~\ref{fig:NN}. Q.X. wrote the section on the method of generation of dataset. Z.Z. provided guidance and funding support. All authors reviewed the manuscript.
    
    \section*{Data availability statement}

    Data is provided within the manuscript or supplementary information files.
    
    \section*{Additional information}

    \textbf{Competing interests} The authors declare no conflicts of interest.
\end{document}